\documentclass{article}

\usepackage[preprint]{neurips_2024}
\usepackage{tikz}
\usetikzlibrary{arrows.meta,positioning,calc}

\usepackage[utf8]{inputenc} 
\usepackage[T1]{fontenc}    
\usepackage[colorlinks, citecolor = blue]{hyperref}       
\usepackage{url}            
\usepackage{booktabs}       
\usepackage{amsfonts}       
\usepackage{nicefrac}       
\usepackage{microtype}      
\usepackage{xcolor}         
\usepackage{amsmath,amssymb,amsthm}
\usepackage{comment}
\usepackage{algorithmicx}
\usepackage{algorithm}
\usepackage{algpseudocode}
\usepackage{bm}
\usepackage{mathrsfs}

\newcommand{\bbeta}{\boldsymbol \beta}
\newcommand{\bepsilon}{\boldsymbol \epsilon}
\newcommand{\bk}{\boldsymbol k}

\newcommand{\bR}{\boldsymbol R}
\newcommand{\bY}{\boldsymbol Y}
\newcommand{\BART}{\operatorname{BART}}

\newcommand{\Categorical}{\operatorname{Categorical}}
\newcommand{\Cor}{\operatorname{Cor}}
\newcommand{\Cov}{\operatorname{Cov}}

\newcommand{\Data}{\mathcal D}
\newcommand{\E}{\mathbb E}
\newcommand{\Exponential}{\operatorname{Exponential}}

\newcommand{\Ghat}{\widehat G}
\newcommand{\GP}{\operatorname{GP}}
\newcommand{\Holder}{Hölder}
\newcommand{\Identity}{\mathrm{I}}
\newcommand{\indep}{\stackrel{\textnormal{indep}}{\sim}}
\newcommand{\iid}{\stackrel{\text{iid}}{\sim}}

\newcommand{\Leaves}{\mathcal L}

\newcommand{\Normal}{\operatorname{Normal}}
\newcommand{\ones}{\mathrm{1}}
\newcommand{\Reals}{\mathbb R}

\newcommand{\sG}{\mathscr G}
\newcommand{\sK}{\mathcal K}
\newcommand{\sH}{\mathcal H}
\newcommand{\sM}{\mathcal M}

\newcommand{\sQ}{\mathcal Q}

\newcommand{\Tree}{\mathcal T}
\newcommand{\Treef}{\operatorname{Tree}}

\newcommand{\Var}{\operatorname{Var}}
\newcommand{\zeros}{\bm 0}

\title{Bayesian Additive Distribution Regression}

\newtheorem{theorem}{Theorem}
\newtheorem{lemma}{Lemma}
\theoremstyle{definition}
\newtheorem{assumption}{Assumption}

\author{
  Antonio R.~Linero$^1$ \And
  Jared Murray$^2$ \And
  Soumyabrata Bose$^1$ \\[0.5em]
  $^1$Department of Statistics and Data Sciences \quad
  $^2$McCombs School of Business \\
  University of Texas at Austin, Austin, TX 78705 \\
  \texttt{antonio.linero@austin.utexas.edu},
  \texttt{sbose@utexas.edu},
  \texttt{jared.murray@mccombs.utexas.edu},
}

\begin{document}

\maketitle

\begin{abstract}
  Distribution regression, where the goal is to predict a scalar response from a distribution-valued predictor, arises naturally in settings where observations are grouped and outcomes depend on group-level characteristics rather than on individual measurements. We introduce DistBART, a Bayesian nonparametric approach to distribution regression that models the regression function as a linear functional with the Riesz representer assigned a Bayesian additive regression trees (BART) prior. We argue that shallow decision tree ensembles encode reasonable inductive biases for tabular data, making them appropriate in settings where the functional depends primarily on low-dimensional marginals of the distributions. We show this both empirically on synthetic and real data and theoretically through an adaptive posterior concentration result. We also establish connections to kernel methods, and use this connection to motivate variants of DistBART that can learn nonlinear functionals. To enable scalability to large datasets, we develop a random-feature approximation that samples trees from the BART prior and reduces inference to sparse Bayesian linear regression, achieving computational efficiency while retaining uncertainty quantification.
\end{abstract}

\section{Introduction}

We consider the problem of distribution regression, given in its simplest form as $Y_i = f(G_i) + \varepsilon_i$ where $Y_1, \ldots, Y_N$ are scalar outcomes, the $\varepsilon_i$'s are mean-$0$ errors, and the $G_i$'s are probability distributions on $\Reals^P$. The aim is to recover $f(\cdot)$ using only samples $X_{ij} \indep G_i$ for $j = 1,\ldots,M_i$. Distribution regression has been applied successfully to many problems, including learning the total mass of dark matter halos \citep{ntampaka2015machine}, classifying text based on bags of word vectors \citep{yoshikawa2014latent}, and addressing the ecological fallacy in estimating political support from population-level data \citep{flaxman2015who}. The unifying feature of all of these problems is that observational units $X_{ij}$ \emph{within} a population (voter characteristics or words in a document) are used to make a prediction about an outcome $Y_i$ defined \emph{at the population level itself}.

In this work, we argue that existing approaches fail to exploit certain desirable structural properties that, when incorporated, improve predictive performance and interpretability. We consider a \emph{sparse additive} structure $f(G_i) = \sum_{v = 1}^V f_v(G_{i,v})$ where $G_{i,v}$ denotes a low-dimensional marginal distribution of $G_i$. In the extreme case, each of the $G_{i,v}$'s is a univariate marginal distribution; this corresponds to $f(G_i)$ depending on $G_i$ only through the marginal distributions of the $X_{ijp}$'s and not (say) on the correlations $\Cor(X_{ij1}, X_{ij2})$ in $G_i$. Similarly, if the $G_{i,v}$'s were bivariate marginals, then $f(G_i)$ might depend on $\Cor(X_{ij1}, X_{ij2})$, but not any higher-order interactions.

We expect a bias towards sparse additive structures to be appropriate in many settings of practical interest. As an example, in social-science applications, group-level outcomes are typically driven by the marginal distributions of a few demographic variables (age, income, education, etc.) and low-order interactions, rather than by the full joint distribution of all predictors. More broadly, the principle that main effects and low-order interactions dominate higher-order ones is a cornerstone of applied statistics \citep{sas2018jmpdoe}. 

We make the following contributions:
\begin{itemize}
\item As a specific method, we propose DistBART, which is based on \emph{Bayesian additive regression trees} (BART, \citealp{ChipmanGeorgeMcCulloch2010}) and possesses these inductive biases. The simplest version of DistBART models $f(G_i)$ as a linear functional $\int \psi(x) \ G_i(dx)$, but we also give an extension to nonlinear functionals. 
\item We empirically demonstrate DistBART's strong performance on both synthetic examples and data from the 2016 U.S. presidential election. 
\item We establish theoretical properties of DistBART, showing that the DistBART posterior contracts at a near-minimax-optimal rate. We also show that DistBART is associated with a kernel mean embedding approach where the kernel is learned from data. 
\item To implement DistBART, we provide both a Gibbs sampling algorithm and a fast approximation that is scalable when the subsample sizes $M_i$ are large. Code reproducing our experiments is given at \url{https://github.com/theodds/DistBART-Replication}.
\end{itemize}

\section{Background}

\subsection{Distribution Regression}

An important feature of distribution regression is that the $G_i$'s are typically not observed and must be estimated from samples $X_{ij} \indep G_i$. Most methods replace $G_i$ with a plug-in estimate $\widehat G_i$, usually the empirical distribution of the $X_{ij}$'s. When the subgroup sample sizes $M_i$ are sufficiently large relative to $N$, there is little lost from doing this. In settings with smaller $M_i$'s, however, it becomes important to take measurement error into account \citep{law2018bayesian}.

\paragraph{Kernel Mean Embeddings}
Many approaches use kernel mean embeddings (KMEs) to first embed the $G_i$'s into a Hilbert space. A comprehensive review of KME-based distribution regression is given by \citet{muandet2017kernel}. Given a bounded kernel function $\kappa : \Reals^P \times \Reals^P \to \Reals$ we can map $G$ into the associated \emph{reproducing kernel Hilbert space} (RKHS) $\sH_\kappa$ through
\begin{align*}
  \phi_G(\cdot) = \int \kappa(x, \cdot) \ G(dx),
\end{align*}
with the RKHS having inner product $\langle f, g \rangle_{\sH_\kappa}$ defined through the reproducing property $\langle \kappa(x, \cdot), f \rangle_{\sH_\kappa} = f(x)$ for all $f \in \sH_\kappa$.

Given such an embedding, it is natural to use kernel-based methods to regress $Y_i$ on $G_i$. \citet{muandet2012learning} use KMEs to construct a support vector machine, while \citet{szabo2016learning} study kernel ridge regression. Concrete examples of kernels include the Gaussian kernel $\sK(\phi_G, \phi_Q) = \exp\{-\lambda \|\phi_G - \phi_Q\|^2_{\sH_\kappa}\}$ and the linear kernel $\sK(\phi_G, \phi_Q) = \langle \phi_G, \phi_Q \rangle_{\sH_\kappa}$. These works focus on two-stage approaches that replace $G_i$ with the empirical distribution $\Ghat_i = \frac{1}{M_i} \sum_{j=1}^{M_i} \delta_{X_{ij}}$ and use the approximate embedding $\phi_{\Ghat_i}(x) = \frac{1}{M_i} \sum_{j=1}^{M_i} \kappa(X_{ij}, x)$ \citep{szabo2015two,bachoc2025improved}. It is also possible to explicitly account for measurement error in the estimated embeddings; \citet{law2018bayesian}, for example, use a Bayesian nonparametric model to sample the $\phi_{G_i}$'s during MCMC-based inference.

\paragraph{Distance-Based Methods} \citet{poczos2013distribution} directly use a kernel smoother $f(G) \approx \frac{\sum_i Y_i \, K\{D(G, \Ghat_i) / h\}}{\sum_i K\{D(G, \Ghat_i) / h\}}$ with $K(\cdot)$ a kernel function and $D(P, Q)$ the total variation distance. The estimates $\Ghat_i$ are kernel density estimates. Rather than using the total variation distance, \citet{meunier2022distribution} take $D(P, Q)$ to be the sliced Wasserstein distance, which can be computed without kernel smoothing and scales more gracefully with the dimension; after computing $D(P,Q)$, they apply kernel ridge regression with (for example) the kernel $\sK(P, Q) = \exp\{-\lambda \, D(P, Q)^2\}$.

\paragraph{Aggregation of Sufficient Statistics}
The problem of aggregating demographic features from individuals to make predictions about populations is also a classical problem in statistics; it appears in the fields of small area estimation \citep{rao2015small}, ecological inference \citep{king2004ecological}, and multilevel modeling \citep{gelman2007data} among others. The $X_{ij}$'s are typically reduced to a set of \emph{sufficient statistics} $\widehat{\phi}_i$, usually consisting of univariate means $\frac{1}{M_i} \sum_{j=1}^{M_i} X_{ij}$ and, possibly, variances. The $\widehat{\phi}_i$'s are treated as noisy proxies for the true population-level parameters $\phi_i$. Shrinkage estimators are then used to account for measurement error \citep{fay1979estimates, battese1988error}. These approaches are limited by their reliance on hand-chosen summaries.

\subsection{Bayesian Additive Regression Trees Priors}

The Bayesian additive regression trees (BART, \citealp{ChipmanGeorgeMcCulloch2010}) framework models an unknown function $\psi(x)$ as a sum of $T$ decision trees
\begin{math}
  \psi(x) = \sum_{t = 1}^T \Treef(x; \Tree_t, \sM_t)
\end{math}
where $\Tree_t$ denotes the structure and decision rules of a binary decision tree and $\sM_t$ denotes the collection of predictions at the leaf nodes. BART uses a \emph{regularization prior} on the tree structures and their leaf parameters that encourages shallow trees with modest contributions. For details on the construction of the prior we use, see the Supplementary Material.

BART can be thought of as a Bayesian version of decision tree boosting \citep{friedman2001greedy}, with the advantage of also giving uncertainty quantification. It has strong empirical and theoretical properties, and is popular among Bayesian statisticians and researchers in causal machine learning. Empirically, BART and its variants \citep{hill2011bayesian, hahn2020bayesian} have consistently performed best in causal inference competitions run by the Society for Causal Inference \citep{dorie2019automated, thal2023causal}. Theoretically, BART optimally estimates the sparse additive structures \citep{rovckova2020posterior,linero2018bayesian}, a property we leverage directly in Section~\ref{sec:adaptive-concentration}.

\section{Tree-Based Additive Decompositions}

\begin{figure}[t]
  \centering \begin{tikzpicture}[
    >=Stealth,
    every node/.style={font=\small}
]

\pgfmathsetmacro{\cx}{1.4}  
\pgfmathsetmacro{\cy}{2.4}  

    
    
    
    


\begin{scope}[shift={(0,0)}]
    \draw[thick, ->] (0,0) -- (4.3,0) node[below, font=\small, pos=1] {$x_1$};
    \draw[thick, ->] (0,0) -- (0,4.3) node[left, font=\small, pos=1] {$x_2$};
    
    \draw (0,0) node[below left, font=\footnotesize] {0};
    \draw (4,0) node[below, font=\footnotesize] {1};
    \draw (0,4) node[left, font=\footnotesize] {1};
    
    \draw[gray!40] (0,0) rectangle (4,4);
    
    
    \begin{scope}
        \clip (0,0) rectangle (0.8,2.4);
        \fill[blue!6] (\cx,\cy) ellipse (2.2 and 1.8);
        \draw[blue!40, thick] (\cx,\cy) ellipse (2.2 and 1.8);
        \draw[blue!55, thick] (\cx,\cy) ellipse (1.5 and 1.2);
        \draw[blue!70, thick] (\cx,\cy) ellipse (0.8 and 0.65);
    \end{scope}
    
    \begin{scope}
        \clip (0.8,0) rectangle (1.8,2.4);
        \fill[blue!6] (\cx,\cy) ellipse (2.2 and 1.8);
        \draw[blue!40, thick] (\cx,\cy) ellipse (2.2 and 1.8);
        \draw[blue!55, thick] (\cx,\cy) ellipse (1.5 and 1.2);
        \draw[blue!70, thick] (\cx,\cy) ellipse (0.8 and 0.65);
    \end{scope}
    
    \begin{scope}
        \clip (0,2.4) rectangle (1.8,4);
        \fill[blue!6] (\cx,\cy) ellipse (2.2 and 1.8);
        \draw[blue!40, thick] (\cx,\cy) ellipse (2.2 and 1.8);
        \draw[blue!55, thick] (\cx,\cy) ellipse (1.5 and 1.2);
        \draw[blue!70, thick] (\cx,\cy) ellipse (0.8 and 0.65);
    \end{scope}
    
    \begin{scope}
        \clip (1.8,0) rectangle (4,1.4);
        \fill[blue!6] (\cx,\cy) ellipse (2.2 and 1.8);
        \draw[blue!40, thick] (\cx,\cy) ellipse (2.2 and 1.8);
        \draw[blue!55, thick] (\cx,\cy) ellipse (1.5 and 1.2);
        \draw[blue!70, thick] (\cx,\cy) ellipse (0.8 and 0.65);
    \end{scope}
    
    \begin{scope}
        \clip (1.8,1.4) rectangle (4,4);
        \fill[blue!6] (\cx,\cy) ellipse (2.2 and 1.8);
        \draw[blue!40, thick] (\cx,\cy) ellipse (2.2 and 1.8);
        \draw[blue!55, thick] (\cx,\cy) ellipse (1.5 and 1.2);
        \draw[blue!70, thick] (\cx,\cy) ellipse (0.8 and 0.65);
    \end{scope}
    
    \begin{scope}
        \clip (0,0) rectangle (4,4);
        \fill[blue!80] (\cx,\cy) circle (2pt);
    \end{scope}
    
    \draw[gray, very thick] (1.8,0) -- (1.8,4);
    \draw[gray, very thick] (0,2.4) -- (1.8,2.4);
    \draw[gray, very thick] (0.8,0) -- (0.8,2.4);
    \draw[gray, very thick] (1.8,1.4) -- (4,1.4);
    
    \node[font=\scriptsize] at (0.4,1.2) {$A_1$};
    \node[font=\scriptsize] at (1.3,1.2) {$A_2$};
    \node[font=\scriptsize] at (0.9,3.2) {$A_3$};
    \node[font=\scriptsize] at (2.9,0.7) {$A_4$};
    \node[font=\scriptsize] at (2.9,2.7) {$A_5$};
    
\end{scope}

\draw[->, thick] (5,2) -- (6.5,2);
\node[align=center] at (5.75,1.4) {Compute\\$G_i(A_\ell)$};

\begin{scope}[shift={(7,0)}]
    
    \node[anchor=west, font=\normalsize] at (0,2) {$\phi_i = $};
    
    \node[anchor=west] at (1.0,2) {%
        $\begin{bmatrix}
            0.11 \\[1pt]
            0.20 \\[1pt]
            0.29 \\[1pt]
            0.05 \\[1pt]
            0.34
        \end{bmatrix}$
    };
    
    \node[anchor=west] at (2.1,2.72) {$= G_i(A_1)$};
    \node[anchor=west] at (2.1,2.36) {$= G_i(A_2)$};
    \node[anchor=west] at (2.1,2.0) {$= G_i(A_3)$};
    \node[anchor=west] at (2.1,1.64) {$= G_i(A_4)$};
    \node[anchor=west] at (2.1,1.28) {$= G_i(A_5)$};
    
    
\end{scope}


\end{tikzpicture}
  \caption{Mapping a spherical Gaussian distribution truncated to $[0,1]^2$,
    $G_i$, to a feature vector $\phi_i$ using trees in the special case where $T
    = 1$.}
  \label{fig:features.tex}
\end{figure}
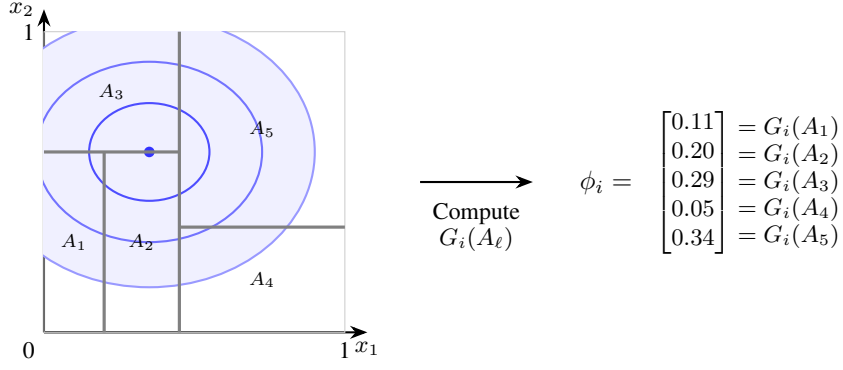

Our starting point is to model $f(G) = \int \psi(x) \ G(dx)$ as a linear functional with Riesz representer $\psi(x)$. This can be interpreted as the effect of nudging $G_i$ in the direction of the point mass $\delta_x$, i.e., $f\{(1-\lambda) G_i + \lambda \delta_x\} = (1 - \lambda) \, f(G_i) + \lambda \, \psi(x)$.  DistBART then models $\psi \sim \BART$ and $\varepsilon_i \sim \Normal(0, \sigma^2)$. Since each tree in the ensemble is a step function over its leaf regions, we can write
\begin{math}
  \psi(x) 
  = \sum_{t=1}^T \sum_{\ell \in \Leaves(\Tree_t)} \mu_{t\ell} \, \ones(x \in A_{t\ell}),
\end{math}
where $A_{t\ell}$ is the region associated to leaf $\ell$ of tree
$\Tree_t$ and $\Leaves(\Tree)$ denotes the set of leaf nodes of $\Tree$. Integrating against $G_i$ gives
\begin{align}
  \label{eq:distbart-features}
  f(G_i) 
  = \sum_{t,\ell} \mu_{t\ell} \, G_i(A_{t\ell}) 
  = \phi_i^\top \bbeta,
\end{align}
where $\phi_i = \{G_i(A_{t\ell})\}$ collects the probabilities that $G_i$ assigns to each region and $\bbeta = \{\mu_{t\ell}\}$ collects the corresponding coefficients. DistBART can therefore be understood as simultaneously learning which features of $G_i$ are relevant and how those features influence the outcome. Figure~\ref{fig:features.tex} illustrates this mapping from a distribution $G_i$ to its feature vector $\phi_i$ for a single tree.

Shallow tree ensembles induce additive decompositions across marginal distributions of $G_i$. If $\Tree_t$ splits only on the variables indexed by $\bk_t \subset \{1, \ldots, P\}$, then each leaf region $A_{t\ell}$ constrains only the coordinates in $\bk_t$, so that $G_i(A_{t\ell})$ depends on $G_i$ only through its $|\bk_t|$-dimensional marginal $G_{i,\bk_t}$. Aggregating across trees, the ensemble decomposes as
\begin{align}
  \label{eq:additive-decomp}
  f(G_i) = \sum_{v=1}^V f_v(G_{i, \bk_v})
\end{align}
where $\bk_1, \ldots, \bk_V$ are the distinct variable subsets appearing across trees.
Crucially, the BART prior places most of its mass on shallow trees with few splits. A tree with a single split on variable $p$ contributes features only for the univariate marginal of $X_{ijp}$; a tree with two splits on variables $p$ and $q$ contributes features depending on the bivariate marginal of $(X_{ijp}, X_{ijq})$; and so forth.

\subsection{Kernel Perspective and Nonlinear Functionals}
\label{sec:kernel-perspective}

Various works have explored connections between tree-based methods and kernel methods \citep{Scornet2016RFKernelMethods, Breiman2000InfinityTheory, BalogLakshminarayananGhahramaniRoyTeh2016MondrianKernel,linero2017review}. DistBART can be connected to kernel-based methods by noting that BART encodes a prior distribution on kernel functions. Let
\begin{math}
  \label{eq:kernel}
  \kappa(x, x') = \frac{1}{T} \sum_{t = 1}^T \kappa_t(x,x') 
\end{math}
where $\kappa_t(x,x')$ is the indicator that $x$ and $x'$ are assigned to the same leaf node of decision tree $t$. As argued by \citet{linero2017review}, if $\psi \sim \BART$ then the limiting distribution as $T \to \infty$ is a Gaussian process with covariance function $\widetilde \kappa(x,x') \propto \E \{\kappa(x,x')\}$ with the expectation being with respect to the prior distribution. \citet{linero2017review} argues that, when the $X_{ij}$'s have uniform marginals, this kernel is approximately $\exp\left\{ -\frac{\lambda \, \|x - x'\|_1}{P} \right\}$ for some $\lambda > 0$, while \citet{petrillo2024gaussian} gives an explicit form for $\widetilde \kappa(x,x')$. We emphasize that our goal is \textbf{not} to approximate this kernel by taking $T$ large, as this would preclude the ability to learn the kernel from data: as noted by \citet{ChipmanGeorgeMcCulloch2010}, there is usually a drop in performance from taking $T$ too large.

Theorem~\ref{thm:connection} establishes that the DistBART model is a type of kernel ridge regression based on a data-adaptive KME. Importantly, the structure of the kernel function $\kappa(x,x')$ is learned from data.

\begin{theorem}
  \label{thm:connection}
  Under the DistBART model, we have $[f \mid \kappa] \sim \GP(0, \sigma^2_\mu \sK)$ where $\sK(G, Q) = \iint \kappa(x,x') \ G(dx) \ Q(dx) = \langle \phi_G, \phi_Q \rangle_{\sH_\kappa}$ and $\sigma^2_\mu / T$ is the prior variance of $\mu_{t\ell}$.
  The posterior mean of $f(\cdot)$ given $\kappa$ is therefore identical to a kernel ridge regression estimator of $f(G)$ under the linear kernel $\langle \phi_G, \phi_Q \rangle_{\sH_\kappa}$.
\end{theorem}

The kernel connection suggests an approach for extending DistBART to nonlinear functionals: we can replace the kernel $\sK(G, Q)$ with, say, a Gaussian kernel $\sK(G, Q) = \exp\{-\gamma \|G - Q\|^2_{\sH_{\kappa}}\}$ so that $f(G)$ would depend only on the marginal distributions determined by the $A_{t\ell}$'s. Alternatively, we can replace the linear modeling layer $Y_i = \phi_i^\top\bbeta + \varepsilon_i$ with a nonlinear model $Y_i = r(\phi_i) + \varepsilon_i$ such that $r(\phi)$ also decomposes additively as $r(\phi_i) = \sum_{v=1}^V r_v(\phi_{i}^{(v)})$ where $\phi_i^{(v)}$ represents a subvector of $\phi_i$; in particular, we will take $r(\phi)$ to be another BART model. 

\subsection{Adaptive Concentration}
  \label{sec:adaptive-concentration}

  We now formalize the claim that DistBART adapts to additive sparse structures. We say that a function $\psi_0: [0,1]^P \to \Reals$ is \emph{$(d, S)$-sparse additive} if there exist index sets $\bk_1, \ldots, \bk_S \subset \{1, \ldots, P\}$ with $|\bk_v| \le d$ for all $v$ and functions $\psi_{0,v}: [0,1]^{|\bk_v|} \to \Reals$ such that $\psi_0(x) = \sum_{v = 1}^S \psi_{0,v}(x_{\bk_v})$. The parameter $d$ controls the maximal order of interaction, while $S$ (assumed less than $T$) controls the number of additive components. We make the following assumptions:

\begin{assumption}[Model]
  \label{ass:model}
  $Y_i = \int \psi_0(x) \, G_i(dx) + \varepsilon_i$ with $\varepsilon_i \iid \Normal(0, 1)$, independently of $\{X_{ij}\}$ and $\{G_i\}$. Additionally, $[X_{kj} \mid \{G_i\}_{i=1}^N] \indep G_k$ and $G_i \iid \sG$ for some meta-distribution $\sG$ on probability distributions on $[0,1]^P$. (Note: the proof strategy can accommodate bounded or $\Normal(0, \sigma^2)$ errors as well.)
\end{assumption}

\begin{assumption}[Prior Thickness]
  For a given rate $\epsilon_N \downarrow 0$ such that $N \, \epsilon_N^2 \to \infty$ there exist positive constants $C_1, C_2$ such that the prior $\Pi(d\psi)$ satisfies $\Pi(\|\psi - \psi_0\|_\infty \le \epsilon_N) \ge C_1 e^{-C_2 \, N \epsilon_N^2}$. Additionally, there exists a $B < \infty$ such that $\|\psi\|_\infty \le B$ with $\Pi$-probability $1$ and $\|\psi_0\|_\infty < B$.
\end{assumption}

\begin{assumption}[Inner Sample Sizes]
  \label{ass:inner}
  We have $\bar M_N^{-1} = o(\epsilon_N^2)$ where $\bar M_N^{-1} = N^{-1} \sum_{i=1}^N M_i^{-1}$.
\end{assumption}

We will study the \emph{fractional posterior}. Let $f_\psi(G) = \int \psi \ dG$ and define the fractional posterior $\Pi^\star_N(A) = \int_A e^{\eta \ell(\psi)} \ d\Pi / \int e^{\eta\ell(\psi)} \ d\Pi$ based on the surrogate likelihood $\ell(\psi) = \sum_i -\frac{(Y_i - f_\psi(\Ghat_i))^2}{2}$ where $\Ghat_i$ is the empirical distribution of the $X_{ij}$'s and $0 < \eta < 1$. The use of fractional posteriors with a surrogate likelihood, as used here, is common in the Safe Bayes \citep{grunwald2012safe} and PAC-Bayes \citep{alquier2024user} literatures, although empirically we have observed that the non-fractional posterior performs similarly. In the Supplementary Material we prove the following result, using techniques inspired by \citet{syring2023gibbs} (see also \citealp{linero2025bayesian} and \citealp{bhattacharya2019bayesian}).

\begin{theorem}
  \label{thm:concentration}
  Under the assumptions above, there exists a constant $U$ depending on $\psi_0$ and the prior such that
  \begin{align*}
    \Pi_N^\star(\|f_\psi - f_{\psi_0}\|_{L_2(\sG)} > U \epsilon_N) = o_P(1),
  \end{align*}
  as $N \to \infty$, where $\|f_\psi - f_{\psi_0}\|^2_{L_2(\sG)} = \int \{f_\psi(G) - f_{\psi_0}(G)\}^2 \ \sG(dG)$.
\end{theorem}

Theorem~\ref{thm:concentration} can be applied to any Bayesian nonparametric prior. The main factor determining the rate is the prior thickness condition, which is where assumptions on the structural form of $\psi_0$ will enter. In the special case of $(d,S)$-sparse additive functions with $\alpha$-\Holder\ smooth components, \citet{rovckova2020posterior} gives conditions for prior-thickness for BART priors with $0 < \alpha \le 1$ at the near-minimax rate $(\log N / N)^{\alpha / (2\alpha + d)}$; \citet{linero2018abayesian} extend this using soft decision trees to $\alpha > 0$. The Supplementary Material gives an application of Theorem~\ref{thm:concentration} with $(d,S)$-sparse additive functions. Beyond this, Theorem~\ref{thm:concentration} shows that the cost of using $\Ghat_i$ in place of $G_i$ is that the rate changes from $\epsilon_N$ to $\max\{\epsilon_N, \bar M_N^{-1/2}\}$.

\paragraph{Remark on Boundedness}
As a technical convenience, we assume $\Pi(\|\psi\|_\infty \le B) = 1$ where $\|\psi_0\|_\infty < B$. A simple way to impose this is to truncate the usual BART prior, i.e., use a modified BART prior with $\Pi_{\text{trunc}}(A) = \Pi(A \cap [\|\psi\|_\infty < B]) / \Pi(\|\psi\|_\infty < B)$. In practice, it is not difficult to specify an upper bound such that truncated and non-truncated BART priors behave the same. We conjecture that the boundedness assumption is not necessary.

\paragraph{Remark on Inner Sample Size}
For the near-minimax rate for $(d,S)$-sparse $\alpha$-\Holder\ smooth functions, Theorem~\ref{thm:concentration} requires $\bar M_N \gg (N / \log N)^{2\alpha / (2\alpha + d)}$. For $\alpha = 1$ and $d = 2$, for example, this is satisfied with $\bar M_N \gg (N/\log N)^{1/2}$. For the voting dataset we have $M_i > N$ for all $i$, and so this condition is comfortably satisfied.

\section{Computation for DistBART}

\paragraph{Fully-Bayesian Inference}
We can sample from the DistBART posterior using a modification of the Gibbs sampling algorithm of \citet{ChipmanGeorgeMcCulloch2010}; the algorithm is given in Algorithm~\ref{alg:bart-dr-backfitting} in the Supplementary Material. This algorithm iteratively updates $(\sigma^2, \{\Tree_t, \sM_t\}_{t = 1}^T)$ with $\Tree_t$ updated via a Metropolis-Hastings transition (using the same GROW, PRUNE, and CHANGE transitions found in the appendix of \citealp{kapelner2016bartmachine}) with $\sM_t$ integrated out.

To perform Gibbs sampling, we first need to obtain the integrated likelihood associated with $\Tree_t$, defined to be $m(\Tree_t) \propto \pi(\Tree_t \mid \{\Tree_k, \sM_k\}_{k \ne t}, \sigma^2)$ where we generically use $\pi(\cdot \mid \cdot)$ to denote a conditional posterior distribution. To do this, we note that
\begin{align*}
  \bY = \sum_t \Phi_t \, \bbeta_t + \bepsilon
\end{align*}
where $\bY = (Y_1, \ldots, Y_N)^\top$, the columns of $\Phi_t$ consist of the derived features $\Ghat_i(A_{t\ell})$, and $\bbeta_t$ consists of the $\mu_{t\ell}$'s. Letting $\bR_t = \bY - \sum_{k \ne t} \Phi_k \, \bbeta_k$, conditional on $\{\Tree_k, \sM_k\}_{k \ne t}$ and $\sigma^2$, this implies the linear model $\bR_t \sim \Normal(\Phi_t \, \bbeta_t, \sigma^2 \Identity)$. Under the prior $\bbeta_t \sim \Normal(\zeros, \sigma^2_\mu \Identity / T)$, the marginal likelihood of $\bR_t$ is $\Normal(\bR_t \mid \zeros, \sigma^2_\mu \, \Phi_t \Phi_t^\top / T + \sigma^2 \, \Identity)$. We therefore have
\begin{align}
  \label{eq:marginal2}
  m(\Tree_t) \propto \Normal(\bR_t \mid \zeros, \sigma^2_\mu \, \Phi_t \Phi_t^\top / T + \sigma^2 \Identity) \times \pi(\Tree_t).
\end{align}
Standard results for Bayesian ridge regression also gives the full conditional $\bbeta_t \sim \Normal\left( \widetilde \bbeta_t, \Sigma_t \right)$ where
\begin{align}
  \label{eq:coefficients}
  \widetilde \bbeta_t = \sigma^{-2} \Sigma_t \Phi_t^\top \bR_t
  \quad \text{and} \quad
  \Sigma_t = (\sigma^{-2} \Phi^\top_t \Phi_t + T \sigma_\mu ^{-2} \, \Identity)^{-1}.
\end{align}
The most computationally expensive part of Gibbs sampling is the computation of the feature matrices $\Phi_t$, which takes $O(T \sum_i M_i)$ time, whereas the remaining computations can be done in $O(TN)$ time using the Woodbury matrix identity.

\paragraph{Fast Random Feature Approximation}
To handle large $M_i$'s, Algorithm~\ref{alg:bart-dr-rf} in the Supplementary Material provides an approximation to fully-Bayesian inference that we found to work well in the regimes we tested. It proceeds by sampling a large number of trees $T$ from the BART prior and computing the features $\phi_{t\ell}(\Ghat_i)$ associated to each $(t, \ell)$. We then fit a linear regression $\bY = \Phi \, \bbeta + \bepsilon$ where $\Phi$ is a matrix containing the $\phi_{t\ell}(\Ghat_i)$'s. To approximate Bayesian inference, we give $\bbeta$ a sparsity-inducing horseshoe prior \citep{carvalho2010horseshoe}; if uncertainty quantification is not desired it suffices to instead use lasso regression \citep{tibshirani1996regression}. We take the number of trees $T$ sampled from the prior to be large, with the sparsity-inducing prior making the choice of tree structures data-adaptive. Similar random feature strategies have been used, for example, by \citet{flaxman2015who,law2018bayesian}.

\section{Model Summarization}

After obtaining posterior samples of $f(G_i)$ or $\psi(x)$, we also require practical tools for interpreting the model. We will use the projection of $\psi(x)$ onto additive models \citep{woody2021model}:
\begin{align*}
  \widetilde \psi(x) = \arg \min_{g \in \sQ} \E_{X_{ij} \sim G} \{g(X_{ij}) - \psi(X_{ij})\}^2
\end{align*}
where $\sQ = \{g : g(x) = \sum_{p = 1}^P g_p(x_p)\}$ is a class of spline-based generalized additive models and $G$ is a reference distribution which can be taken to be (a subset of) the empirical distribution of the full set of $X_{ij}$'s. The functions $g_p(x_p)$ can then be interpreted as an additive effect due to $X_{ijp}$. We can also model the quality of this approximation using the \emph{summary-$R^2$} defined by
\begin{align*}
  R^2 = 1 - \frac{\E_{X_{ij} \sim G}\{\psi(X_{ij}) - \widetilde \psi(X_{ij})\}^2}
                 {\E_{X_{ij} \sim G} \{\psi(X_{ij}) - \bar \psi\}^2}
  \quad \text{where} \quad
  \bar \psi = \E_{X_{ij} \sim G} \psi(X_{ij}).
\end{align*}
Summary-$R^2$ can be used in combination with the projection $\widetilde \psi(x)$ to understand both the additive contributions of the variables and to understand the degree to which $\psi(x)$ is driven by interactions (with a low summary-$R^2$ indicating that interactions between variables are important).

To interpret the overall importance of the features we use a leave-one-covariate-out (LOCO, \citealp{rinaldo2019bootstrapping}) variable importance. This approach first splits the data into training and evaluation sets, and then fits $P + 1$ models to the training data: the full model, and a model with each covariate removed. We then compute the test-set $R^2$ of predictions vs. outcomes with variable $p$ removed. The variable importance is then defined as the reduction in $R^2$ from omitting variable $p$ relative to the $R^2$ of the full model.

\section{Experiments}

\subsection{Synthetic Data}
\label{sec:synthetic-data}

We now use synthetic data to test the hypothesis that DistBART will provide better predictions than KME-based approaches when the underlying data has a sparse additive structure.

\paragraph{Data Generating Mechanisms}
We vary the data generating mechanism across the following variables: (i) the generative model for the $G_i$'s, (ii) the functional form of $f(G)$, (iii) the number of sampled distributions $N \in \{100, 200, 400, 800, 1600\}$, and (iv) the number of predictors $P \in \{5, 10, 20, 40\}$. To model the $G_i$'s, we take the $X_{ijp}$'s to be marginally exponentially distributed (with mean sampled from an $\Exponential(1)$ distribution) or normally distributed (with mean sampled from a $\Normal(0,1)$ distribution); these marginals were linked together via a Gaussian copula with a uniformly-sampled correlation matrix. For $f(G)$ we consider both a \emph{sparse setting} with $\psi(x) = x_1 \, x_2 + x_3 \, x_4$ and a \emph{main effects setting} $\psi(x) = \sum_{p=1}^P x_p$.

\paragraph{Benchmark Methods}
We consider tree-based (DistBART) features, kernel mean embeddings with a Gaussian kernel (RBF) features, mean (Mean) features that summarize each group by the marginal means $\bar X_i = \frac{1}{M_i}\sum_j X_{ij}$, and a hybrid approach (Both) that uses both RBF and BART features. The lasso was used after featurizing the $G_i$'s for all methods to obtain predictions. In the case of the Gaussian kernel features, following \citet{law2018bayesian}, landmarks were chosen using $K$-means clustering. We note that the Mean method exactly captures the main effects setting, and so a priori should perform well under that setting. All settings used $M_i = 200$ for all groups, which is smaller than the $M_i$'s for the voting dataset.

\paragraph{Metrics}
We evaluate predictive performance using root mean-squared error (RMSE) in estimating $f(G_i)$, which was averaged over $10$ replications at each simulation setting. Mean-squared error on a given replication was computed as $\frac{1}{N} \sum_i \{f(G_i) - \widehat f(G_i)\}^2$ where $\widehat f(G_i)$ denotes the predicted value for distribution $G_i$.

\begin{figure}
  \centering
  \includegraphics[width=1\textwidth]{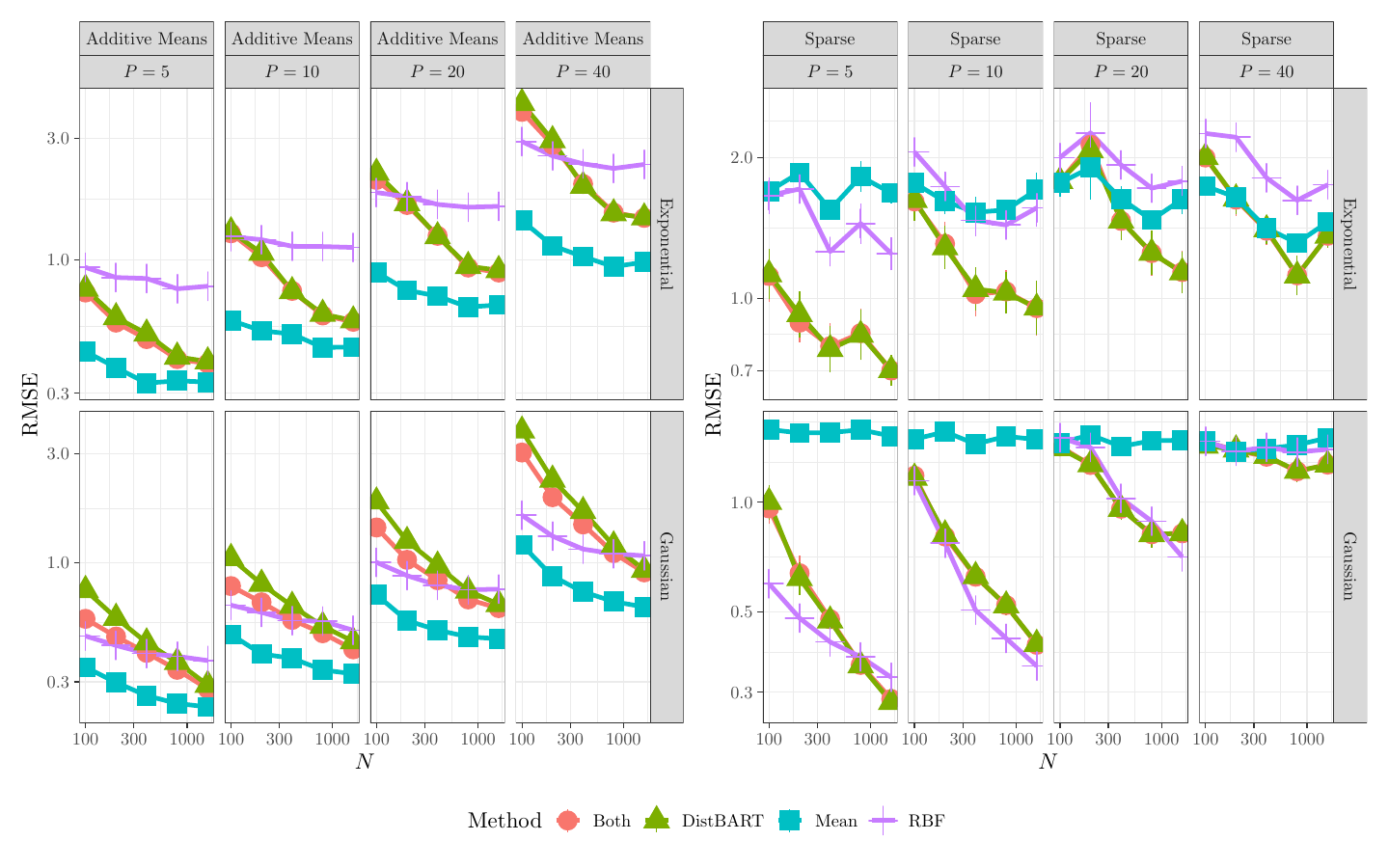}
  \caption{
    Mean test RMSE for four methods: BART-based features, RBF-based features, both BART and RBF-based features, and mean features, aggregated using \texttt{glmnet} for the simulation experiment described in Section~\ref{sec:synthetic-data}.}
  \label{fig:simulation_experiment}
\end{figure}

\paragraph{Results}
Figure~\ref{fig:simulation_experiment} displays average RMSE as a function of sample size $N$. When the data generating mechanism is exponential, we see that BART gives a sizable improvement over RBF features, with the hybrid approach performing very similarly. Interestingly, when the $G_i$'s are multivariate normal, we find that the RBF features perform well. We conjecture that this occurs due to the fact that RBFs with a Gaussian kernel are well-adapted to estimate smooth functionals of multivariate Gaussian distributions, and for small $N$ this outweighs the benefits of DistBART adapting to additive structures. In the case of exponential marginals, however, the RBF performs poorly. Predictably, for the main effects settings the mean features perform best, while mean features perform poorly in the sparse setting. All methods become worse as $P$ increases.

\subsection{Voting Data}
\label{sec:voting}

\begin{figure}[t]
  \centering
  \includegraphics[width=0.8\textwidth]{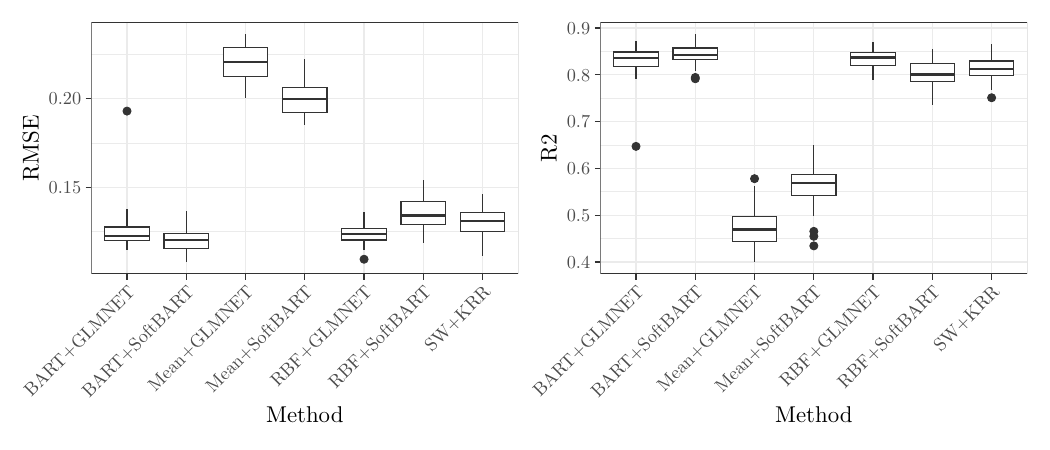}
  \caption{Performance comparison of distribution regression on the voting dataset, based on 30 repeated 80-20 train/test splits.}
  \label{fig:method-comparison}
\end{figure}

We next use DistBART to analyze results from the 2016 U.S. presidential election. The dataset consists of individual-level observations from the American Community Survey aggregated to Public Use Microdata Areas (PUMAs), with the outcome being the vote gap $Y_i = (D_i - R_i) / T_i$ where $D_i$, $R_i$, and $T_i$ denote Democratic, Republican, and total votes, respectively, within PUMA $i$. For each PUMA we observe a sample of individuals characterized by seven demographic features: age, sex, race, household income (which we log-transform), employment status, citizenship status, and education status. Distribution regression is required here because individual-level demographics are available, but the outcome of interest --- aggregate voting behavior --- is observed only at the population level. The number of individuals included in the data is $\sum_i M_i = 9,861,010$ ($\min M_i \approx 2000$) across $N = 979$ PUMAs. We compare seven methods, varying both the PUMA-level features and the prediction model. We consider BART features, Gaussian kernel embeddings (RBF), and marginal mean features. We consider both the lasso tuned by cross-validation and the soft BART model of \citet{linero2018abayesian} as downstream prediction models. We tuned the depth parameter $\beta$ of the BART prior as well as the bandwidth of the Gaussian kernel embeddings. We also included a sliced Wasserstein kernel ridge regression \citep{meunier2022distribution} using the kernel $K(P, Q) = \exp\left\{ -\gamma D(P,Q)^2 \right\}$ with $\gamma$ and the shrinkage parameter $\lambda$ tuned by cross-validation as an additional non-linear non-tree-based baseline. Figure~\ref{fig:method-comparison} reports results based on 30 splits of the data into training and testing splits, evaluated using RMSE and the squared correlation between predictions and outcomes on test data ($R^2$) across the different splits.

First, we see that the mean features performed much worse than all competing methods, confirming that higher-order distribution information beyond first moments is important for this problem. Beyond this, all methods perform reasonable well, with the best performing method being the non-linear variant of the DistBART procedure followed by the linear versions of both the Gaussian kernel emebddings and DistBART.

\begin{figure}[t]
  \centering
  \includegraphics[width=1\textwidth]{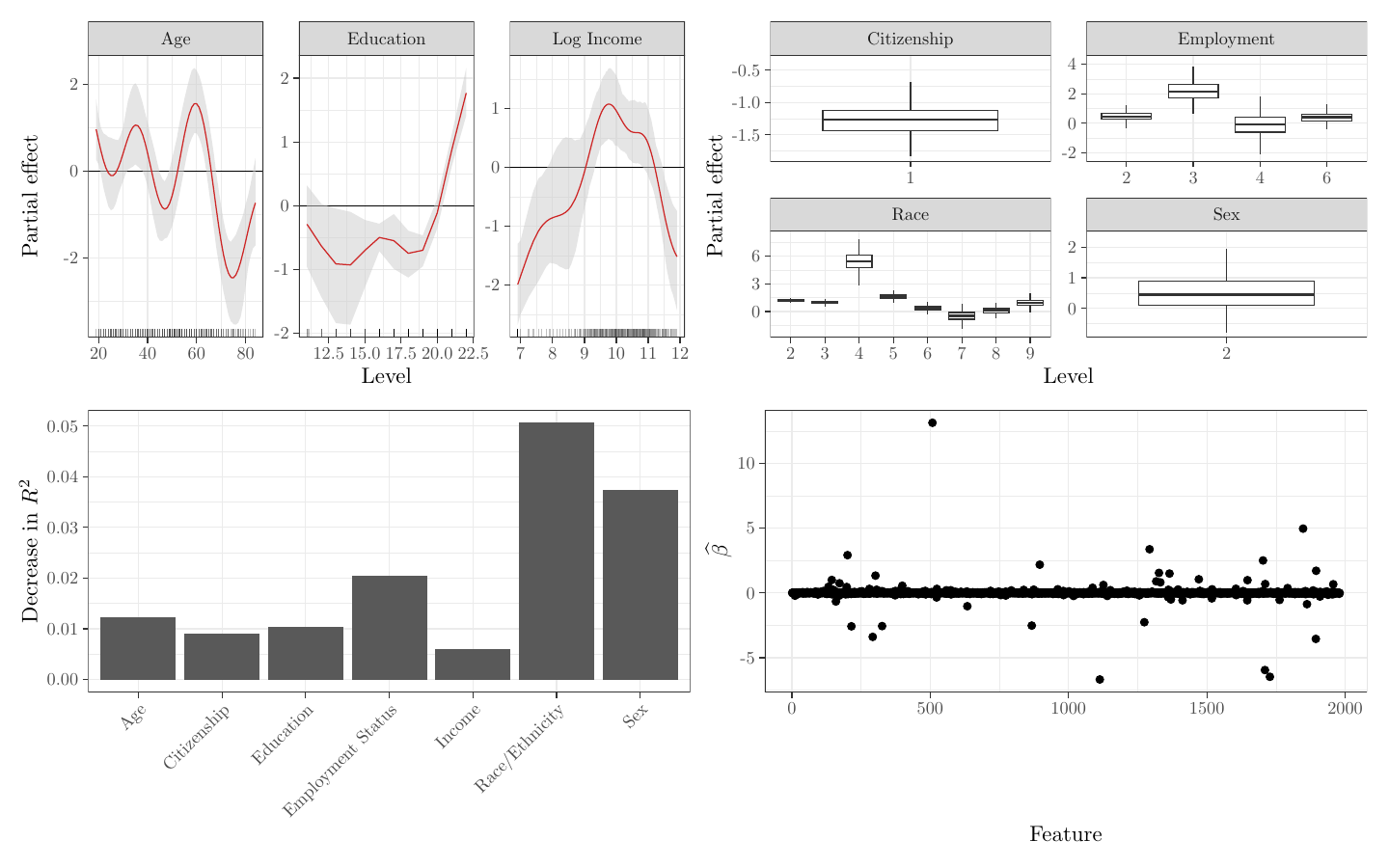}
  \caption{
    Posterior summaries of the estimated linear functional from the voting analysis. Top: posterior mean and 95\% credible bands for additive summaries of $\psi(x)$ over continuous covariates (left) and categorical (right) covariates. Bottom left: Reduction in test-set $R^2$ when each feature is omitted. Bottom right: posterior mean of feature coefficients from a horseshoe regression fit on the tree-derived features. Codes for categorical features are given in the Supplementary Material.
  }
  \label{fig:voting}
\end{figure}

Our final analysis model combined the BART features with a horseshoe regression for the outcome. Additive model summaries are given in Figure~\ref{fig:voting}. We see that the summaries for all three continuous variables are nonlinear; for example, increasing education of a population beyond high school sharply increases Democratic vote-share, while the effect of income is non-monotone, with both low-income and top-income populations shifting the vote share towards Republicans. In the case of the effect of age, we suspect that the evident nonlinearity is a symptom of the additive model being inadequate, as the quality of the additive summary is quite poor (summary-$R^2 \approx 30\%$). This agrees with the fact that the demographic factors that impact voting have strong interaction effects.

Figure~\ref{fig:voting} also displays LOCO variable importances. LOCO ranks race distribution highest in terms of importance, followed by sex, employment status, and age. Lastly, Figure~\ref{fig:voting} gives the estimated regression coefficients from the horseshoe. The most important feature encoded an interaction between age and sex, suggesting that this is an important interaction accounted for by DistBART.


\section{Discussion}
\label{sec:discussion}


\paragraph{Limitations}
The motivation for using DistBART comes from an assumption that $f(G_i)$ decomposes additively across low-dimensional marginals. While we believe this is often the case for tabular datasets, there are settings where this assumption is inappropriate. If the features $X_{ij}$ were images represented as pixel intensities, it is implausible that $f(G_i)$ would depend primarily on marginal distributions of individual pixels rather than on their spatial relationships. In such settings, methods that first transform the $X_{ij}$'s nonlinearly into meaningful embeddings before aggregating to the distribution level would be more effective. Additionally, we did not attempt to account for measurement error in this work, as it is less important in the $M_i > N$ regime. In some other settings, such as predicting school-level outcomes, we have $N \gg M_i$ and measurement error becomes much more important; we plan to pursue this in future work.

\paragraph{Extensions}
Beyond distribution regression, our approach can be applied to correlated random effects models for hierarchical data \citep{mundlak1978pooling,bell2015explaining}, where we observe individual-level outcomes nested within groups. The goal shifts to modeling how both individual characteristics $X_{ij}$ and the distribution of characteristics within the group $G_i$ jointly affect outcomes $Y_{ij}$. By constructing tree-based features and including them as group-level predictors in a mixed effects model, we can flexibly capture contextual effects --- the influence of group composition on individual outcomes --- while allowing for correlation between group-level random effects and within-group covariate distributions. This provides a nonparametric alternative to classical correlated random effects models that rely on group means. This will be pursued in future work.

\bibliographystyle{apalike}
\bibliography{references.bib}

\clearpage

\appendix

\section{Proof of Theorem~\ref{thm:connection}}

Fix the tree ensemble (equivalently, fix $\kappa$). Recall that
\begin{align*}
  \psi(x)=\sum_{t=1}^T\sum_{\ell\in\Leaves(\Tree_t)} \mu_{t\ell}\,\ones(x\in A_{t\ell}),
  \qquad
  \mu_{t\ell} \indep \Normal(0, \sigma^2_\mu / T).
\end{align*}
So, for any distribution $G$,
\begin{math}
  f(G)=\int \psi(x)\,G(dx)
  =\sum_{t=1}^T\sum_{\ell\in\Leaves(\Tree_t)} \mu_{t\ell}\,G(A_{t\ell}).
\end{math}
As $f(G)$ and $f(Q)$ are different linear combinations of the same Gaussian random variables, $[f\mid \kappa]$ is mean-zero Gaussian with covariance
\begin{align*}
  \Cov\{f(G),f(Q)\mid \kappa\}
  &=\sum_{t=1}^T\sum_{\ell\in\Leaves(\Tree_t)} \Var(\mu_{t\ell})\,G(A_{t\ell})\,Q(A_{t\ell}) \\
  &=\frac{\sigma_\mu^2}{T}\sum_{t=1}^T\sum_{\ell\in\Leaves(\Tree_t)} G(A_{t\ell})\,Q(A_{t\ell}).
\end{align*}
Recalling $\kappa_t(x,x') = \sum_{\ell \in \Leaves(\Tree_t)} 1(x \in A_{t\ell}) \, 1(x' \in A_{t\ell})$, we also have
\[
  \iint \kappa_t(x,x')\,G(dx)\,Q(dx')
  = \sum_{\ell\in\Leaves(\Tree_t)} G(A_{t\ell})\,Q(A_{t\ell}).
\]
Therefore,
\begin{align*}
  \Cov\{f(G),f(Q)\mid \kappa\}
  &=\frac{\sigma_\mu^2}{T}\sum_{t=1}^T \iint \kappa_t(x,x')\,G(dx)\,Q(dx') \\
  &=\sigma_\mu^2 \iint \kappa(x,x')\,G(dx)\,Q(dx')
  =\sigma_\mu^2\,\sK(G,Q),
\end{align*}
which proves $[f\mid \kappa]\sim \GP(0,\sigma_\mu^2\sK)$.
Next, since $\phi_G(\cdot)=\int \kappa(x,\cdot)\,G(dx)$, the reproducing property gives
\[
  \langle \phi_G,\phi_Q\rangle_{\sH_\kappa}
  =\iint \langle \kappa(x,\cdot),\kappa(x',\cdot)\rangle_{\sH_\kappa}\,G(dx)\,Q(dx')
  =\iint \kappa(x,x')\,G(dx)\,Q(dx')
  =\sK(G,Q).
\]
Thus the posterior mean under the GP prior is given by $[f\mid\kappa]\sim \GP(0,\sigma_\mu^2\sK)$ with $\sK(G, Q) = \langle  \phi_G, \phi_Q \rangle_{\sH_\kappa}$.

Finally, suppose $Y_i=f(G_i)+\varepsilon_i$ with $\varepsilon_i\iid\Normal(0,\sigma^2)$. Let $K\in\Reals^{N\times N}$ have entries $K_{ij}=\sigma_\mu^2\sK(G_i,G_j)$ and let $k(G)\in\Reals^N$ have entries $k_i(G)=\sigma_\mu^2\sK(G,G_i)$. By standard Gaussian conditioning,
\[
  \E\{f(G)\mid \kappa,\Data\}
  =
  k(G)^\top (K+\sigma^2\Identity)^{-1}\bY
  \quad \text{where} \quad \bY = (Y_1, \ldots, Y_N)^\top.
\]
Equivalently, writing $\widetilde K_{ij}=\sK(G_i,G_j)$ and $\widetilde k_i(G)=\sK(G,G_i)$,
\[
  \E\{f(G)\mid \kappa,\Data\}
  =
  \widetilde k(G)^\top\!\left(\widetilde K+\frac{\sigma^2}{\sigma_\mu^2}\Identity\right)^{\!-1}\bY,
\]
which is the kernel ridge regression predictor with kernel $\sK(G,Q)$ and ridge parameter $\lambda=\sigma^2/\sigma_\mu^2$. Since $\sK(G,Q)=\langle \phi_G,\phi_Q\rangle_{\sH_\kappa}$, this is kernel ridge regression with the linear kernel on the KME features.


\section{Proof of Theorem~\ref{thm:concentration}}

We now prove our posterior contraction result. Recall that we assume $G_1, G_2, \ldots \iid \sG$, $X_{ij} \indep G_i$ given $\{G_i\}$, and $Y_i = f_{\psi_0}(G_i) + \varepsilon_i$ where $\varepsilon_i \iid \Normal(0, 1)$ independently of all other quantities. Here, we have adopted the notation $f_\psi(G_i) = \int \psi(x) \ G_i(dx)$. Let
\begin{align*}
  \|f_\psi - f_{\psi_0}\|^2_{L_2(\sG)} = \E_{G_i \sim \sG}\{f_\psi(G_i) - f_{\psi_0}(G_i)\}^2.
\end{align*}
be the squared $L_2(\sG)$ distance and define the surrogate log-likelihood by
\begin{align*}
  \ell(\psi) = \frac{-\sum_{i = 1}^N \{Y_i - f_{\psi}(\Ghat_i)\}^2}{2}.
\end{align*}

Given a prior $\Pi(d\psi)$, the fractional posterior based on the pseudo-likelihood is given by
\begin{align*}
  \Pi^\star_N(A) = \frac{\int_A R(\psi) \ d\Pi}{\int R(\psi) \ d\Pi}
  \quad \text{where} \quad R(\psi) = e^{\eta\{\ell(\psi) - \ell(\psi_0)\}}.
\end{align*}
Now, we define sets $A^\star = [\|f_\psi - f_{\psi_0}\|^2 \le \epsilon_N^2]$ and $A = [\|f_\psi - f_{\psi_0}\|^2 \le U^2 \epsilon_N^2]$ for a constant $U$ that will be taken large later; the goal is to show that $\Pi^\star_N(A^c) \to 0$ in probability, with the set $A^\star$ used as part of a technical device to control the denominator.

Throughout, we let $\E_0$ and $\Var_0$ denote the expectation and variance with respect to the true distribution of $\{Y_i, G_i, X_{ij}\}$.

We make the following assumptions:
\begin{description}
\item[Prior Thickness] The prior satisfies $\Pi\{\|f_\psi - f_{\psi_0}\|_{L_2(\sG)} \le \epsilon_N\} \ge C_1 e^{-C_2 N \epsilon_N^2}$ for positive constants $C_1$ and $C_2$.
\item[Boundedness] There exists a $B < \infty$ such that $\|\psi\|_\infty \le B$ holds $\Pi$-almost surely. Additionally, $\|\psi_0\|_\infty < B$.
\item[Inner Sample Size] The inner sample sizes grow sufficiently fast in the sense that $\bar M_N^{-1} = o(\epsilon_N^2)$.
\end{description}
The only technical condition to check is the prior thickness condition, which fortunately can be easily verified for many priors (including BART priors) used in practice. For example, for the target rate $\epsilon_N = N^{-\alpha / (2 \alpha + d)}$ when $\psi_0 \in C^{\alpha}_B([0,1]^P)$ is $(d,S)$-sparse additive, there are multiple works that verify the stronger requirement $\Pi\{\|\psi - \psi_0\|_\infty \le \epsilon_N\} \ge C_1 e^{-C_2 \, N \epsilon_N^2}$ \citep{rovckova2020posterior,linero2018abayesian}. 

\subsection{Control of the Denominator}

We begin by computing a high-probability lower bound for the denominator. We first have
\begin{align}
  \label{eq:lb1}
  \int R(\psi) \ d\Pi
  \ge \int_{A^\star} R(\psi) \ d\Pi
  = \Pi(A^\star) \int R(\psi) \ d\Pi_{A^\star}
\end{align}
where $\Pi_{A^\star}$ denote the restriction of $\Pi$ to the set $A^\star$. Define the random variable
\begin{align*}
  Z = \int \{\ell(\psi) - \ell(\psi_0)\} \ d\Pi_{A^\star}.
\end{align*}
By Jensen's inequality, we can lower bound \eqref{eq:lb1} by 
\begin{align*}
  \Pi(A^\star) \exp(\eta Z) \ge \Pi(A^\star) \exp(-\eta|\bar Z| - \eta|Z - \bar Z|)
\end{align*}
where $\bar Z = \E_0(Z)$. By Fubini's theorem,
\begin{align*}
  \bar Z = \int \E_0\{\ell(\psi) - \ell(\psi_0)\} \ d\Pi_{A^\star}.
\end{align*}
Direct calculation of $\E_0\{\ell(\psi) - \ell(\psi_0)\}$ gives
\begin{align*}
  \sum_i \frac{-\|f_\psi - f_{\psi_0}\|^2_{L_2(\sG)}}{2} -\frac{\E_0 \Var_0(f_\psi(\Ghat_i) \mid G_i)}{2}
  + \frac{\E_0 \Var_0(f_{\psi_0}(\Ghat_i) \mid G_i)}{2}.
\end{align*}
Computing the variance gives
\begin{align}
  \label{eq:varb}
  \Var_0(f_\psi(\Ghat_i) \mid G_i) = \frac{\Var_0\{\psi(X_{i1}) \mid G_i\}}{M_i} \le \frac{B^2}{M_i}
\end{align}
with the inequality following from the boundedness assumption $\|\psi\|_\infty \le B$.  Applying the triangle inequality and the definition of $A^\star$, we have
\begin{align*}
  |\bar Z| \le \frac{N \epsilon_N^2}{2} + \frac{B^2 N}{\bar M_N}.
\end{align*}
To finish controlling the denominator we require a high-probability upper bound for $|Z - \bar Z|$. By Chebyshev's inequality, we have $|Z - \bar Z| \le N \, \epsilon_N^2 + B^2 N / \bar M_N$ holds with probability at least
\begin{math}
  1 - \frac{\Var_0(Z)}{(N \, \epsilon_N^2 + B^2 N / \bar M_N)^2}.
\end{math}
Define
\begin{align*}
  V_i(\psi, \psi_0) = \E_0 \left\{ -\frac{(Y_i - f_{\psi}(\Ghat_i))^2 - (Y_i - f_{\psi_0}(\Ghat_i))^2}{2} \right\}^2,
\end{align*}
and observe that by Jensen's inequality we have
\begin{align*}
  \Var_0(Z)
  &=
  \sum_i \Var_0 \int \left\{ -\frac{(Y_i - f_{\psi}(\Ghat_i))^2 - (Y_i - f_{\psi_0}(\Ghat_i))^2}{2} \right\} \ d\Pi_{A^\star}
  \\&\le
    \sum_i \E_0 \left[ \int \left\{ -\frac{(Y_i - f_{\psi}(\Ghat_i))^2 - (Y_i - f_{\psi_0}(\Ghat_i))^2}{2} \right\} \ d\Pi_{A^\star} \right]^2
  \\&\le
  \sum_i \int V_i(\psi, \psi_0) \ d\Pi_{A^\star}.
\end{align*}
Our goal will be to show that $V_i(\psi, \psi_0) \le C(B) \{\|f_\psi - f_{\psi_0}\|^2_{L_2(\sG)} + B^2 / M_i\}$ for some constant $C(B)$, which would ensure that the bound holds with probability at least $1 - \frac{C(B)}{N \epsilon_N^2 + B^2 N / \bar M_N}$. To do this, we temporarily adopt the shorthand $a = (f_{\psi_0}(G_i) - f_\psi(G_i))$, $b = (f_\psi(G_i) - f_{\psi}(\Ghat_i))$, and $c = (f_{\psi_0}(G_i) - f_{\psi_0}(\Ghat_i))$. The numerator term in $V_i(\psi, \psi_0)$ can be rewritten as
\begin{align*}
  (\varepsilon_i + a + b)^2 - (\varepsilon_i + c)^2
  = a^2 + b^2 - c^2 + 2a \varepsilon_i + 2b \varepsilon_i - 2c \varepsilon_i.
\end{align*}
By Cauchy-Schwarz, the square of this is bounded by
\begin{align*}
  6(a^4 + b^4 + c^4 + 4 a^2 \varepsilon_i^2 + 4b^2 \varepsilon_i^2 + 4c^2 \varepsilon_i^2).
\end{align*}
Taking the expectation and using the bound $a^2, b^2, c^2 \le 4B^2$ gives the further bound
\begin{align*}
  6(4B^2 + 4) (\E_0(a^2) + \E_0(b^2) + \E_0(c^2))
  \le 24 (B^2 + 1)(\|f_{\psi} - f_{\psi_0}\|^2_{L_2(\sG)} + \frac{2 B^2}{M_i}),
\end{align*}
using the fact that, as in \eqref{eq:varb}, $\E_0(b^2)$ and $\E_0(c^2)$ are bounded by $B^2 / M_i$. We therefore have, on $A^\star$,
\begin{align*}
  V_i(\psi, \psi_0) \le 48(B^2 + 1)\{\|f_{\psi} - f_{\psi_0}\|^2_{L_2(\sG)} + \frac{B^2}{M_i}\} \le
  48(B^2 + 1) \{\epsilon^2_N + B^2 / M_i\}.
\end{align*}
We therefore conclude that $|Z - \bar Z| \le N \epsilon_N^2 + B^2 N / \bar M_N$ occurs with probability at least $1 - \frac{48(B^2 + 1)}{(N\epsilon_N^2 + B^2 N / \bar M_N)} = 1 - o(1)$. Applying the prior thickness assumption, this finally gives that the denominator is lower-bounded by
\begin{align*}
  C_1 \exp\left\{ -(2\eta + C_2)N \epsilon_N^2 - 2\eta B^2 N / \bar M_N  \right\}
\end{align*}
on a set $\Lambda$ with probability $1 - o(1)$.

\subsection{Control of the Numerator}

We apply Fubini's theorem to get
\begin{align*}
  \E_0 \int_{A^c} R(\psi) \ d\Pi
  = \int_{A^c} \E_0 R(\psi) \ d\Pi
  = \int_{A^c} \E_0 e^{\eta\{\ell(\psi) - \ell(\psi_0)\}} \ d\Pi
  = \int_{A^c} \prod_{i = 1}^N  \E_0 e^{\eta\{\ell_i(\psi) - \ell_i(\psi_0)\}} \ d\Pi,
\end{align*}
where $\ell_i(\psi) = -\frac{(Y_i - f_{\psi}(\Ghat_i))^2}{2}$. For the purpose of this section, we temporarily let $a = f_{\psi}(\Ghat_i) - f_{\psi_0}(\Ghat_i)$ and $b = f_{\psi_0}(G_i) - f_{\psi_0}(\Ghat_i)$. Algebra then gives
\begin{align*}
  \ell_i(\psi) - \ell_i(\psi_0) = -\frac{a^2}{2} + a(b + \varepsilon_i).
\end{align*}
Using the moment generating function of a standard normal and taking the expectation conditional on $\{G_i, X_{ij}\}$, we have
\begin{align*}
  \E_0 e^{\eta\{\ell_i(\psi) - \ell_i(\psi_0)\}} =
  \E_0 \exp\left\{ -\frac{\eta (1 - \eta) a^2}{2} + \eta \, a \, b \right\}.
\end{align*}
Let $c_0 = \eta(1 - \eta) / 2 > 0$. By $2 uv \le u^2 + v^2$ with $u = \sqrt{c_0}\,a$ and $v = \eta b/\sqrt{c_0}$, we have
\begin{align*}
  \E_0 e^{\eta\{\ell_i(\psi) - \ell_i(\psi_0)\}} 
  &\le
  \E_0 \left\{ \exp\left( -\frac{c_0 a^2}{2} \right) \exp\left( \frac{\eta^2 b^2}{2c_0} \right) \right\},
  \\&\le 
  \left\{ \E_0 \exp\left( -c_0 a^2 \right)\right\}^{1/2} \left\{\E_0 \exp\left( \frac{\eta^2 b^2}{c_0} \right) \right\}^{1/2},
\end{align*}
where the second line follows from Cauchy-Schwarz. We now bound the two terms separately using the following lemma, which we prove later.

\begin{lemma}
  \label{lem:expb}
  If $W$ is supported on $[0,A]$ with $\E_0(W) \le V$ and $s \ge 0$ then $\E_0 e^{sW} \le \exp\left\{ C(s, A) V \right\}$ for the constant $C(s,A) = \frac{e^{sA} - 1}{A}$. The same bound also holds for $s < 0$ if we instead have a lower bound $\E_0(W) \ge V$.
\end{lemma}

\paragraph{Bound for $b$} Applying Lemma~\ref{lem:expb} to $b^2$, we know from \eqref{eq:varb} that $\E_0(b^2) \le \frac{B^2}{M_i}$ and that $b^2$ is supported on $[0,4B^2]$, so taking $W = b^2$, $s = \frac{\eta^2}{c_0}$, and $V = B^2 / M_i$, and $A = 4B^2$, we obtain
\begin{align*}
  \left\{\E_0 \exp\left( \frac{\eta^2 b^2}{c_0} \right) \right\}^{1/2}
  \le
  \exp\left( K(\eta, B) \frac{B^2}{M_i} \right)
\end{align*}
for some constant $K(\eta, B)$.

\paragraph{Bound for $a$} We again apply Lemma~\ref{lem:expb} with
\begin{align*}
  \E_0(a^2)
  &= \E_0 \{f_\psi(\Ghat_i) - f_{\psi_0}(\Ghat_i)\}^2
  \\&= \E_0 [\Var\{f_\psi(\Ghat_i) - f_{\psi_0}(\Ghat_i) \mid G_i\} + \{f_{\psi}(G_i) - f_{\psi_0}(G_i)\}^2]
  \\&\ge \|f_{\psi} - f_{\psi_0}\|^2_{L_2(\sG)}.
\end{align*}
Hence, for some $K_0(\eta, B)$ we have on the set $A^c = [\|f_\psi - f_{\psi_0}\|^2_{L_2(\sG)} > U^2 \epsilon_N^2]$ that
\begin{align*}
  \left\{\E_0 e^{-c_0 a^2}  \right\}^{1/2}
  \le \exp\left\{ -K_0(\eta, B) \|f_\psi - f_{\psi_0}\|^2_{L_2(\sG)} \right\}
  \le \exp\left\{ -K_0(\eta, B) U^2 \epsilon_N^2 \right\}.
\end{align*}

\paragraph{Putting the Bounds Together} Applying the bounds for $a$ and $b$, we have
\begin{align*}
  \E_0 \int_{A^c} R(\psi) \ d\Pi
  \le \prod_i \exp\left\{ -K_0(\eta, B) U^2 \epsilon_N^2 + K(\eta, B) B^2 / M_i \right\}.
\end{align*}
By Markov's inequality, we therefore have
\begin{align*}
  \int_{A^c} R(\psi) d\Pi = O_P\left( \exp\left\{ -K_0(\eta, B) U^2 N \epsilon_N^2 + K(\eta, B) B^2 N / \bar M_N \right\} \right).
\end{align*}

\subsection{Controlling the Posterior}

Let $\Lambda = [|Z - \bar Z| \le N \epsilon_N^2 + B^2 N / \bar M_N]$. Then
\begin{align*}
  \Pi^\star_N(A^c)
  &=
  \frac{\int_{A^c} R(\psi) \ d\Pi}{\int R(\psi) \ d\Pi}
  \\&\le
  \frac{\int_{A^c} R(\psi) \ d\Pi}{C_1 \exp\{-(2\eta + C_2)(N \epsilon_N^2 - 2 \eta B^2 N / \bar M_N)\}} + 1(\Lambda^c)
  \\&=
  \frac{O_P\left( \exp\left\{ -K_0(\eta, B) U^2 N \epsilon_N^2 + K(\eta, B) N B^2 / \bar M_N \right\} \right)}{C_1 \exp\{-(2\eta + C_2)(N \epsilon_N^2 - 2 \eta B^2 N / \bar M_N)\}} + o_P(1)
\end{align*}
By taking $U$ sufficiently large and recalling the assumption that $\bar M^{-1}_N = o(\epsilon_N^{2})$, the dominating term of the ratio is $e^{-K_0(\eta, B) U^2 N \epsilon_N^2}$ and so $\Pi_N^\star(A^c) = o_P(1)$.

\subsection{Proof of Lemma~\ref{lem:expb}}

Using convexity of the exponential function $f(w) = e^{sw}$ we have for $w \in [0,A]$ that
\begin{align*}
  e^{sw} \le 1 + \frac{e^{sA} - 1}{A} w.
\end{align*}
Taking the expected value gives
\begin{align*}
  \E_0 e^{sW}
  \le 
  1 + \frac{e^{sA} - 1}{A} \E_0(W) 
  \le
  \exp\left\{ \frac{e^{sA} - 1}{A} \E_0(W) \right\}.
\end{align*}
If $s > 0$, then the term $\frac{e^{sA} - 1}{A}$ is positive, so setting $\E_0(W) \le V$ gives an upper bound, while if $s < 0$ then the term $\frac{e^{sA} - 1}{A}$ is negative, so setting $\E_0(W) \ge V$ gives an upper bound.


\section{Algorithms}

This section collects pseudocode for the computational procedures used in the paper. Algorithm~\ref{alg:bart-dr-backfitting} presents the fully Bayesian backfitting (Gibbs) sampler for DistBART, which iteratively updates the tree ensemble and associated parameters. Algorithm~\ref{alg:bart-dr-rf} describes the random-tree feature approximation used for scalable inference, in which trees are sampled from the prior to construct distribution-level features followed by a downstream regression model. The random features algorithm is particularly useful when $\sum_i M_i$ is large.

\begin{algorithm}[t]
\caption{Bayesian backfitting (Gibbs) sampler for DistBART}
\label{alg:bart-dr-backfitting}
\begin{algorithmic}[1]
  \Require Grouped data $\{(X_{ij})_{j=1}^{M_i}, Y_i\}_{i=1}^N$, a BART prior, and a transition kernel $Q(\Tree \mid \Tree')$ on the decision trees, number of MCMC iterations $B$ to run
  \Ensure Posterior draws of $\{(\Tree_t,\sM_t)\}_{t=1}^T$ and $\sigma^2$; fitted values $f(G_i)$.
  \State Initialize parameters by sampling from the prior and compute $\phi_{i,(t,\ell)} = \frac{1}{M_i} \sum_{j = 1}^{M_i} 1(X_{ij} \in A_{t\ell})$ for all $i,t,\ell$.
\For{iteration $b=1,\ldots,B$}
  \For{$t=1,\ldots,T$}
  \State $R_i \gets Y_i - \sum_{k \ne t} \sum_{\ell \in \Leaves(\Tree_k)} \phi_{i, (k,\ell)} \mu_{k\ell}$
  \State Propose $\Tree^\star \sim Q(\Tree \mid \Tree_t)$
  \State Compute the integrated likelihoods $m(\Tree_t)$ and $m(\Tree^\star)$ given in \eqref{eq:marginal2}
  \State Set $\Tree_t \gets \Tree^\star$ with probability $\min\left\{\frac{m(\Tree^\star) \, Q(\Tree_t \mid \Tree^\star)}{m(\Tree_t) \, Q(\Tree^\star \mid \Tree_t)}, 1\right\}$; recompute $\phi_{i, (t,\ell)}$'s if accepted
  \State Sample $\sM_t$ from the conditional \eqref{eq:coefficients}
  \EndFor
  \State Sample $\sigma^2$ from its full conditional distribution
  \EndFor
\end{algorithmic}
\end{algorithm}

\begin{algorithm}[t]
\caption{Random-tree features for distribution regression}
\label{alg:bart-dr-rf}
\begin{algorithmic}[1]
\Require Grouped data $\{(X_{ij})_{j=1}^{M_i}, Y_i\}_{i=1}^N$, and a BART prior
\Ensure Posterior for $f(G_i)$
\For{$t=1,\ldots,T$}
  \State Sample $\Tree_t$ from the prior, with $\{A_{t\ell}\}_{\ell=1}^{L_t}$ the leaf regions of tree $t$ for $L_t$ leaves
\EndFor
\State For all $(i,t,\ell)$, set $\phi_{i,(t,\ell)} \gets \frac{1}{M_i}\sum_{j=1}^{M_i} 1(X_{ij}\in A_{t\ell})$
\State Create design matrix $\Phi \in \Reals^{N \times \sum_t L_t}$ from the $\phi_{i,(t,\ell)}$'s, dropping constant columns 
\State Fit a horseshoe regression model $Y_i = \phi_i^\top \, \bbeta + \varepsilon_i$ or soft BART $Y_i = r(\phi_i) + \varepsilon_i$ model and return samples $f(G_i) = \phi_i^\top\bbeta$ or $f(G_i) = r(\phi_i)$ accordingly.
\end{algorithmic}
\end{algorithm}

\section{Details on the BART Prior}
\label{sec:details}

The prior over trees is defined as follows. A node at depth $d$ is non-terminal with probability $\alpha (1 + d)^{-\beta}$, where we use the common default values $\alpha = 0.95$ and $\beta = 2$ recommended by \citet{ChipmanGeorgeMcCulloch2010}, which concentrates the prior on trees with between $1$ and $5$ leaf nodes (probabilities 0.05, 0.55, 0.28, 0.09, and 0.03 respectively).

Given the tree topology, decision rules of the form $[x_j \le c]$ are sampled as follows:
\begin{enumerate}
\item A splitting index $j$ is sampled according to a $\Categorical(s)$ distribution where $s = (s_1, \ldots, s_P)$ are prior probabilities of selecting index $j$.
\item A split point $c$ is sampled uniformly from $[A_j, B_j]$ where $\prod_{p = 1}^P [A_p, B_p]$ defines the hyperrectangle of $x$ values associated to the current node.
\end{enumerate}
As a preprocessing step, we transform all of the $X_{ijp}$'s to lie in the interval $[0,1]$. To do this, we compute the empirical distribution of each of the $X_{ijp}$'s across all groups and then map each $X_{ijp}$ to its associated percentile. For categorical variables we use a one-hot encoding.

Given the decision tree, the leaf node value $\mu_{t\ell} \in \sM_t$ associated to leaf $\ell$ of $\Tree_t$ is given a Gaussian prior $\mu_{t\ell} \sim \Normal(0, \sigma^2_\mu)$, with $\sigma^2_\mu \propto T^{-1}$. We write $r \sim \BART(T, \alpha, \beta, \sigma_\mu)$ (or just $r \sim \BART$ when convenient) to denote that $r(\cdot)$ has a BART prior with $T$ trees, tree parameters $(\alpha, \beta)$, and leaf node variance $\sigma^2_\mu$.

\section{Application of Theorem~\ref{thm:concentration} to BART}

Below, we give conditions on $\psi_0(x)$ and the prior $\Pi(d\psi)$ such that Assumption 2 of Theorem~\ref{thm:concentration} holds. A proof that these conditions are sufficient is given by \citet{linero2018abayesian}; it uses \emph{soft} decision trees in place of standard decision trees, and so allows for $\alpha$-\Holder\ smooth functions for all $\alpha > 0$, however the same result remains true for $0 < \alpha \le 1$ if standard decision trees are used instead.

A generic ``smooth'' decision tree is written in the form
\begin{align*}
  \psi(x) = \sum_{t, \ell \in \Leaves(\Tree_t)} \mu_{t\ell} \, \varpi_{t\ell}(x)
\end{align*}
where
\begin{align*}
  \varpi_{t\ell}(x) = \prod_{b \in \mathcal A(t,\ell)} F\left(\frac{x_j - c_b}{\tau} \right)^{L_b(x)} \left[ 1 - F\left(\frac{x_j - c_b}{\tau}  \right) \right]^{1 - L_b(x)}
\end{align*}
where $L_b(x)$ is the indicator that the path from the root node to the leaf node $\ell$ goes ``left'' at branch $b$, $\mathcal A(t,\ell)$ denotes the set of \emph{ancestor nodes} (i.e., the nodes on the path from the root to the leaf) associated to leaf $\ell$ of tree $t$, and $F(\eta) = (1 + e^{-\eta})^{-1}$. The parameter $c_b$ plays the role of the splitting point and $\tau$ is a smoothness parameter that, in the limit $\tau \to 0$, leads to a hard decision.

We make the following assumptions:

\begin{enumerate}
\item[A1] The function $\psi_0: [0,1]^P \to \Reals$ is $\alpha$-\Holder\ smooth for some $\alpha > 0$ and is $(d,S)$-sparse.
\item[A2] There are some constants $(C_1, C_2)$ such that the prior distribution on the number of trees $T$ satisfies $\Pi(T = t) \geq C_1 \exp(-C_2 t)$ for $t = 0, 1, 2, \ldots$ 
\item[A3] The prior density $\pi_\tau$ on the bandwidth parameter $\tau$ satisfies $\pi_\tau(\tau) \geq a_1 \tau^{a_2}$ for some constants $a_1, a_2 > 0$ for all sufficiently small $\tau$.
\item[A4] The prior on the splitting proportion vector $s$ is $\operatorname{Dirichlet}(a/p^\xi, \ldots, a/p^\xi)$ for some $\xi > 1$ and $a > 0$.
\item[A5] The leaf coefficients $\mu_{t\ell}$ are independent and identically distributed with density $\pi_\mu$ where $\pi_\mu(\mu) \geq B_1 \exp(-B_2 |\mu|)$ for all $\mu$ and some positive constants $B_1$ and $B_2$.
\item[A6] $\Pi(D_t = k) > 0$ for $k = 0, 1, \ldots, 2d$, where $D_t$ denotes the depth of tree $t$.
\end{enumerate}

All other details in prior are the same as those in Section~\ref{sec:details}; the main modification required is that the leaf node parameters are sub-exponential rather than Gaussian. Under these assumptions, \citet{linero2018abayesian} prove that there are positive constants $A, C$ such that
\begin{align*}
  \Pi\left( \|\psi - \psi_0\|_\infty \le A \, \epsilon_N \right)
  \ge e^{-C \, N \, \epsilon_N^2}
\end{align*}
where $\epsilon_N = N^{-\alpha / (2\alpha + d)} (\log N)^t + \sqrt{\frac{d \log P}{N}}$ for any $t \ge \alpha (d + 1) / (2\alpha + d)$. In addition to allowing for $(d,S)$-sparsity, this bound also allows for high-dimensional covariates with the additional term $\sqrt{\frac{d\log P}{N}}$ determining how fast $P$ can diverge (note that the number of variables that $\psi_0(x)$ depends on cannot diverge to apply this result).

\section{Codes for Categorical Features}

\textbf{Race:}
    \begin{enumerate}
        \item White alone
        \item Black or African American alone
        \item American Indian alone
        \item Alaska Native alone
        \item American Indian and Alaska Native tribes specified; or American Indian or Alaska Native, not specified and no other races
        \item Asian alone
        \item Native Hawaiian and Other Pacific Islander alone
        \item Some Other Race alone
        \item Two or More Races
    \end{enumerate}

\textbf{Citizenship:} 1 = Born in the U.S.; 0 = otherwise.

\textbf{Sex:} 1 = Male; 2 = Female.

\textbf{Employment Status:}
    \begin{enumerate}
        \setcounter{enumii}{0}
        \item[N/A] Less than 16 years old
        \item Civilian employed, at work
        \item Civilian employed, with a job but not at work
        \item Unemployed
        \item Armed forces, at work
        \item Armed forces, with a job but not at work
        \item Not in labor force
    \end{enumerate}

\end{document}